# Dynamics of Size-Selected Gold Nanoparticles Studied by Ultrafast Electron Nanocrystallography

*Chong-Yu Ruan\*, Yoshie Murooka, Ramani K. Raman, Ryan A. Murdick*

Department of Physics and Astronomy

Michigan State University, East Lansing, MI 48824, USA

\* corresponding author: ruan@pa.msu.edu

ABSTRACT We report the studies of ultrafast electron nanocrystallography on size-selected Au nanoparticles (2-20 nm) supported on a molecular interface. Reversible surface melting, melting, and recrystallization were investigated with dynamical full-profile radial distribution functions determined with sub-picosecond and picometer accuracies. In an ultrafast photoinduced melting, the nanoparticles are driven to a non-equilibrium transformation, characterized by the initial lattice deformations, nonequilibrium electron-phonon coupling, and upon melting, the collective bonding and debonding, transforming nanocrystals into shelled nanoliquids. The displasive structural excitation at premelting and the coherent transformation with crystal/liquid coexistence during photomelting differ from the reciprocal behavior of recrystallization, where a hot lattice forms from liquid and then thermally contracts. The degree of structural change and the thermodynamics of melting are found to depend on the size of nanoparticle.

Understanding the phases of materials and their transformations is a fundamental problem, especially on the nanometer scale, where thermodynamics and kinetic processes are influenced by local environments, including surfaces, microsctructures, and interfacial chemistry[1,2]. Their elucidation requires characterization at the atomistic scale, both in space and time; at nanointerfaces molecular sensitivity is necessary. The possibility for a scattering experiment to couple with the high time resolution of a femtosecond laser in a pump-probe arrangement makes it a favorable option to study structural dynamics, as evident from recent developments[3-6]. Here we report a nanocrystallographic method, based on Ultrafast Electron Crystallography (UEC)[7], which allows quantitative studies of local structures and transient dynamics of nanoparticles (NPs) dispersed on a molecular interface. The implementation is general. Specifically demonstrated here are the studies of reversible photoinduced melting and subsequent recrystallization of size-selective Au NPs (2nm, 10nm, 20nm), a prototypical system for studying nanophases[1] and catalysis[2], supported on a molecular surface. The size-dependent phase transitions are examined using a more bulk-like NP (20nm, melting point ~ 1300K[8]) and a much smaller one, where surface and confinement play significant roles (2nm, melting point ~ 800K[8]). By achieving spatial and thermal energy isolation of NPs from their environment and from each other, the normally irreversible phase transformations become reversible, allowing multi-shot pump-probe diffraction to map out their full courses. Such implementation allows the use of a low-density electron pulse to avoid the pulse-broadening effect[9] and has high data reproducibility compared with single-shot experiments where a much higher density electron pulse is required[10].

The spatial and thermal isolation of the NPs from their environment is achieved by implementing a buffer molecular layer, in this case aminosilane, self-assembled on a silicon substrate as shown in Fig. 1A [11,12], with which, substrate scattering is sufficiently suppressed. Since the NPs are dispersed, the diffraction is via transmission, mostly through individual particles rather than multiple particles (or aggregates). To highlight the difference, two cases are presented in Fig. 1 (D-I). Without buffering, NPs tend to aggregate, as visible in the electron micrograph (D). The diffraction pattern (E) is dominated by the substrate, as also evident from the rocking curve (F). With buffering, however, NPs are separated from each other on the surface (G), from which the diffraction (H) is predominantly from the NPs and the buffering molecular layer, with no indication of Si periodicity in the rocking curve (I). Samples were also examined following the laser irradiation experiment and showed no signs of agglomeration or damage. A time-resolved structural study of surface supported Au NPs was conducted earlier using a synchrotron X-ray source[13]. However, because X-ray is much more penetrating, with 5 orders of magnitude less scattering power than electron, its interfacial structural resolution is limited by the signal-to-noise level, particularly for very small NPs. High-energy electron diffraction, with its short wavelength and high surface sensitivity as demonstrated here, has higher structural resolution than small angle X-ray diffraction.

Prior to studying dynamics, the static structure of NPs is analyzed. The static pattern shows Debye-Scherrer diffraction rings from the Au NPs, while the ordered buffer layer produces Bragg spots, primarily in the surface streak regions. The Debye-Scherrer rings are radially averaged into a 1D diffraction intensity curve, shown in Fig. 2A. This curve, obtained from the diffraction pattern of 2nm NPs, shows isolated peaks, reflecting a crystalline structure, which, from inspection, resembles more towards cuboctahedral and/or decahedral structures than an icosahedral one. Here, $s = (4\pi/\lambda)\sin(\theta/2)$ represents the magnitude of the reciprocal space wave-vector of diffracted electrons, with wavelength $\lambda=0.069$ Å for 30keV electrons, and $\theta$ being the electron scattering angle. The deviation from an ideal structure can be attributed to different possible conformations and the surface strain associated with small particles[14]. Their cuboctahedra-like crystalline characteristics are more easily seen in the modified radial distribution functions (mRDF)[15], deduced from a Fourier analysis of 1D diffraction curves, shown in Fig. 2B. All the major peaks above 2.8Å are in close agreement with the Au-Au distance table based on a face-center cubic (FCC) motif (Fig. 2C), which constitutes the internal lattice repeat of a cuboctahedra. The sensitivity of our technique is sufficient to permit the observation of molecular density peaks as well in the mRDF, such as those at 1.5 - 1.7 Å representing the C-C, C-N, and Si-N bonds and ~1.1 Å for C-H and N-H bonds.

To study the dynamics, an ultrashort laser pulse (800 nm, P-polarized, ~40 fs) is used to excite the NPs, while the probing electron pulse is delayed relative to the laser pulse to monitor the structural evolution (Fig. 1B). To improve temporal resolution, a proximity-coupled optical system allows the photogenerated electron beam to be focused to ~5 μm in less than 6 cm from the photocathode with 1000 or less electrons/pulse in order to remediate the space charge induced broadening effects and to reduce the pulse overlap between the pump and probe. Sub-ps accuracy can be readily achieved (Fig. 1C). Using different excitation fluences (tuned to nonmelting, surface melting, and melting), transient responses of atoms in the NPs are determined from dynamical full profile mRDFs, highlighting the local dynamics, compared with the global ones based on analyzing Bragg peaks. The dynamics of bonds following laser irradiation can be extracted from the mRDF maps, shown in Fig. 3, selected here with the surface melting (31 mJ/cm$^2$) and melting (75 mJ/cm$^2$) fluences for 2nm NP. Their differentiation is evident from the rapid change in bond densities. In general, a melting is characterized by the replacement of sharp 2$^{nd}$ nearest neighbor peaks with more diffusive ones[7]. Uniquely here, the peak density reduction is coupled to the formation of new density peaks at slightly larger distances. This redistribution of 2$^{nd}$ nearest neighbor peaks can be used to determine the onset of melting as well as recrystallization, defined here as a *1/e drop in peak intensity* at ~5Å. Based on this criterion, at 31 mJ/cm$^2$, we observe no melting. The laser heating causes the NPs to expand with little adjustment of bond densities below 1nm. At 18 ps, breaking and forming of bonds beyond 1nm are evident, indicative of surface melting. The transition is rapid, within 1-2 ps, and the molten layer lasts for only tens of ps. By 40 ps the newly formed long bonds begin to cool and slowly replace the original broken ones. The surfaces revert to original crystalline structure in ns timescale. However, at 75 mJ/cm$^2$, the bond densities across the full NP length scale are modified and complete melting occurs. Using the change in 2$^{nd}$ nearest neighbor peak, we determine the occurrence of melting and recrystallization at 18 ps and 110 ps respectively. Furthermore, the melting and recrystallization dynamics display vastly different characters, nonreciprocal to each other, as seen in the mRDF map. Following the initial expansion of the lattice, we can clearly see bonding and debonding emerge already at ~ 12 ps. The depletion of the bond density (debonding) around the major peaks (numbered 3,5,7,12 in Fig. 2B) is coupled to the enhancement of bond density (bonding) at longer distances, where bonds may or may not have been present before. The emerging longer distance peaks, which constitute a shoulder region, gain in density towards melting, ultimately smearing out into bands. The coherent bonding and debonding dynamics observed at the premelting period and the continued development of the newly formed long distance peaks into liquid structures suggests that liquid structures are populated while the particle is still relatively cold. This reflects the displasive character of the photo-melting process during which the transformation of crystal into liquid is through breaking old bonds and forming new bonds. In a sense, this photomelting dynamics resembles a conformation change between minimum energy structures on the free energy landscape. In contrast, the recrystallization is much like the reverse of a 'thermal' melting in which crystal simply thermally expands and then disorders. The liquid structure of a NP is unique, and can be characterized by the reduced coordination number, judging from the reduction of the direct bond density, and shell-like mRDF densities. The structure of the liquid is compared with an icosahedral NP, which also possesses shell-like structure. The average distances of the liquid shells in the mRDF, taken at 40 ps, match well with those of a 10% expanded icosahedral shells with the pronounced crystalline peaks being smeared out. This suggests the density of the transient hot nanoliquid is reduced compared with a room temperature crystalline structure and the atom-atom correlation between liquid shells is lost.

Closer inspection of the NPs expansion before melting reveals anisotropic movement of the lattice depending on the irradiating fluence. By fitting the mRDF density profiles using Gaussian function, we follow the time-dependent evolutions of the bond distance, density, and width for bonds at 2.88, 5.00, and 7.64 Å. At a low fluence (<31 mJ/cm$^2$), the changes are isotropic with thermal expansion being equal for all three distances. However, the deformation of the lattice sets in as the fluence increases. At a threshold fluence of 38 mJ/cm$^2$, right before a full melting occurs, the early time (1-15 ps) anisotropic bond movements are evident, representing a combination of shearing motion along (100) direction and expansion (Fig. 4A, left panel). Generally, prior to lattice disorder, the nearest neighbor bond (2.88 Å) sharpens while the rest of the longer bonds decay and widen. The transient narrowing at direct bond distance suggests a brief reduction of strain in the NP following the impulsive laser excitation. Expansion of direct bond continues till 60 ps, indicating uninterrupted transfer of energy into bond stretching vibrations. However, these slower dynamics do not represent the time scale of electron-lattice equilibration. The intrinsic electron-phonon coupling time for Au NPs is less than 4±1 ps, a limit derived based on fitting the Debye-Waller factor from low fluence (15 mJ/cm$^2$) data where no lattice disorder or coherent motion is evidently present at short times. Thus the longer period for melting (18-20 ps) and the lattice expansion (60 ps) reflect the time scales for atomic disorder in the crystal and thermal energy relaxation from the initially strongly excited hot phonons to the bond stretching vibrations. Link and El-Sayed[16] have found time constant of 30 ps for NP shape change from nanorod to nanosphere, a time scale comparable to phonon-phonon scattering time.

Melting of Au NPs has been investigated by other time-resolved techniques also. Plech and coworkers have reported the melting transition of Au NPs (100nm) using synchrotron based time-resolved X-ray powder diffraction, on the 100 ps time scale (their pulse duration)[17]. The sample was irradiated by a 400 nm fs laser with steadily increasing power, and the phase changes were interpreted based on monitoring the deviation of the integrated area under (111) and (200) Bragg peaks from a constant value, at a fixed delay of 105 ps. The corresponding lattice temperature is derived based on the lattice expansion by monitoring the shift of Bragg peak position. They observe a sub-bulk melting temperature (70% of the bulk value), which is unexpected for particle size larger than 30 nm. They attribute this suppression of melting temperature to possible onset of surface melting. The largest lattice expansion before melting was determined to be 1.2% (1.82% is expected for bulk melting), and there was no indication of any significant crystal anisotropy based on diffraction. In a more recent small-angle X-ray scattering (SAXS) study, Plech and coworkers found 15 mJ/cm$^2$ as the melting threshold for 38 nm Au NPs, and have shown laser alignment effects just below the melting fluence[18]. Hartland, Hu, and Sader have addressed the melting transition by measuring the vibration frequency of the breathing mode using time-resolved spectroscopy, but have found no discontinuity at the melting point[19]. They conclude that a saturation of light absorption limits the energy that can be transferred to the lattice. To connect our data to these studies, we also inspect the temporal evolution of Bragg peaks in *s*-space. At 38 mJ/cm$^2$ (the threshold fluence for 2nm NPs), we find a rapid decay of intensity (~6 ps for (111) peak to drop by *1/e*). Such a significant change, however, does not correspond to a melting, as shown from our mRDF analyses, rather it indicates a breaking of lattice symmetry induced by photo-excitation. This deformation is also evident from the anisotropic shifts of different lattice planes – (220) blue-shifts while (311) and (331) peaks red-shift. Their associated peak-widths exhibit instantaneous narrowing followed by widening, again confirming the coherent change at short times. These shifts are consistent with a lattice deformation along (100) direction. The lattice then expands significantly to a maximum of ~1.5% at 60 ps. The 'lattice' is found significantly disordered, no longer suitable for s-space analyses. However, the disorder is just below the threshold considered as melting according to our mRDF analyses. At 75 mJ/cm$^2$ (melting fluence), a rapid drop of (111) intensity to 20% of the original level is found to appear at 15 ps, indicating the rapid loss of long-range order. This time scale is close to the onset of bonding and debonding observed in the premelting period according to the mRDF analyses. For 20nm Au NPs, we found the threshold fluence to be between 15-20 mJ/cm$^2$, consistent with the results for 38 nm NPs obtained by Plech and coworkers.

To understand the thermal energy redistribution, we use the stretch of bonds to gauge the 'local temperature' of bonds in NPs. The term 'local' used here suggests a temperature based on the vibrational sampling of local bonding potential between atoms. The anharmonicity of the bonding potential leads to the expansion of the bond, which increases with the vibrational amplitude. Coherent motion resulting from impulsive strain at early times from the fs excitation will cause splitting or broadening (if unresolved) of peaks or driven anisotropic deformation of the lattice. To this end, coarse-graining the dynamics with longer time period should be conducted to reflect the average extension of the lattice due to thermal (stochastic) energy. This method based on the mRDF analysis allows differentiation of inhomogeneity that exists on different length scales, not possible by extracting temperature solely based on following the shift of a Bragg peak [17]. However, such a definition of temperature must not be confused with the temperature of the NP as a whole, which can only be defined when the thermal equilibrium is reached. By using the long-time data, where thermalization has been established, we can extract the NPs' true temperatures under different fluences by comparing them to a two-temperature model (TTM) [20] (Fig. 4A). To convert the thermal expansion into temperature, we use the temperature-dependent thermal expansion coefficient from reference 21 which is valid between 300 and 1300 K. This thermal expansion coefficient is found to apply for 60nm Au NPs[17]. Comparing the long-time thermal relaxation, which is fit to a TTM, with the short-time heightened lattice expansion reveals the hot phonons effect caused by non-equilibrium electron-phonon coupling (Fig. 4A, right panel). The existence of hot phonons was recently invoked to explain the non-equilibrium electron-phonon coupling in low-dimensional systems, such as graphite[22] and nanotube[23], as well as molecular systems[24]. They are the vibrational modes coupled more directly to the de-excitation of electrons, thus gaining higher 'temperature' compared with an equilibrated lattice temperature obtained from a TTM. These hot phonons produce large amplitude of vibration, thus leading to heightened lattice expansion. Because the phonon-phonon interaction time is 30-60 ps, these hot phonons are likely responsible for initiating melting and influencing recrystallization, making the photomelting phenomena different from a thermal one. Size-dependent effects in the transient heating of NPs are seen, shown in Fig. 4B. First, the transient maximum bond stretch is significantly higher in the 2nm NPs at 75 mJ/cm$^2$ (6% for 2nm, 3.5% for 20nm), albeit, the thermal temperature is very close in both cases - a fact deduced by comparing the equilibrated (ΔR/R) data at longer times (from 1-3 ns data, see insets). Second, 20nm NPs have similar maximum bond stretch at 80 mJ/cm$^2$ and 31 mJ/cm$^2$, both leading to melting, but differing in their liquid residence time. For 2nm NPs however, melting occurs only at 75 mJ/cm$^2$. These results suggest that the particle size plays a role in determining the thermodynamics of NPs. For 20 nm NPs, increasing the fluence does not cause a continuous rise in liquid temperature, leading instead to a longer liquid residence time, suggesting that latent heat already exists at 20 nm. The lack of a sharp transition expected from first-order phase transition reflects the ultrafast nature of transformation. However, for 2nm NPs, the temperature continues to rise significantly after melting, suggesting the transformation being a second-order phase transition[25,26].

Based on the TTM[27], at the irradiating fluence of 31 mJ/cm$^2$, 1.5×10$^{22}$ e$^-$/cm$^3$ (~24%) are excited in the Au NP, whereas at 75 mJ/cm$^2$, 3.25×10$^{22}$ e$^-$/cm$^3$ (~57%) are excited. At these high fluences, the interband transition starts to play a role. The lowest interband transition energy in Au is 1.7 eV[28], which corresponds to promoting *d*-electrons (*5d*) in the vicinity of the *X*-point of the first Brillouin zone to the conduction band *(6sp)* and is slightly higher than our excitation energy (1.55 eV). However, as conduction electrons are strongly excited, their Fermi-Dirac distribution is modified, with part of the electronic levels below the Fermi level being emptied, to make way for interband transition. Because of this hot electron effect, the contribution of the interband transition increases with fluence. The effect of interband transition is manifested in the lattice anisotropic deformation observed at the short times (1-15 ps). Although, the *d* holes relaxation will proceed in tens of fs by hole-hole scattering[29], the lattice deformation likely persist to the ps time scale due to the slow collective motion of atoms responding to the modification of the energy landscape caused by the core electron (*5d*) excitation. The coupling of ps lattice deformation to electronic heating in bulk system was also discussed recently by Guo and Taylor[30]. In addition, we find that by using 400 nm excitation this anisotropic deformation is replaced by an isotropic one. Because of the high excitation energy, the interband transition is no longer pinned to the X-point. These results suggest that the excited energy landscape can be explored by following the ultrafast lattice dynamics as a function of the excitation energy.

In conclusion, using ultrafast electron nanocrystallography, we have mapped out the dynamics of liquid-crystalline and crystalline-liquid phase transformations for Au nanoparticles, at and beyond the thermodynamic limit. The accurate mRDF determinations of nanostructures allow quantitative studies of atomic dynamics with molecular scale resolutions. The size dependence is evident in the change of structures and in the extent of melting. The reversible and coherent transformation on the ultrafast time scale demonstrates the directed dynamics on the energy landscape of finite systems. Abundant details can be further extracted by comparing the dynamical mRDF maps and electron micrographs obtained for fluence far beyond the melting threshold, and under different excitation wavelengths. This methodology is general and could be implemented to study a wide class of phenomena pertaining to nanoscaled materials.

ACKNOWLEDGMENT

This work is supported by the US Department of Energy, Office of Basic Energy Sciences, Division of Material Sciences and Engineering and the Intramural Research Grant Program at Michigan State University.

**Figure 1**

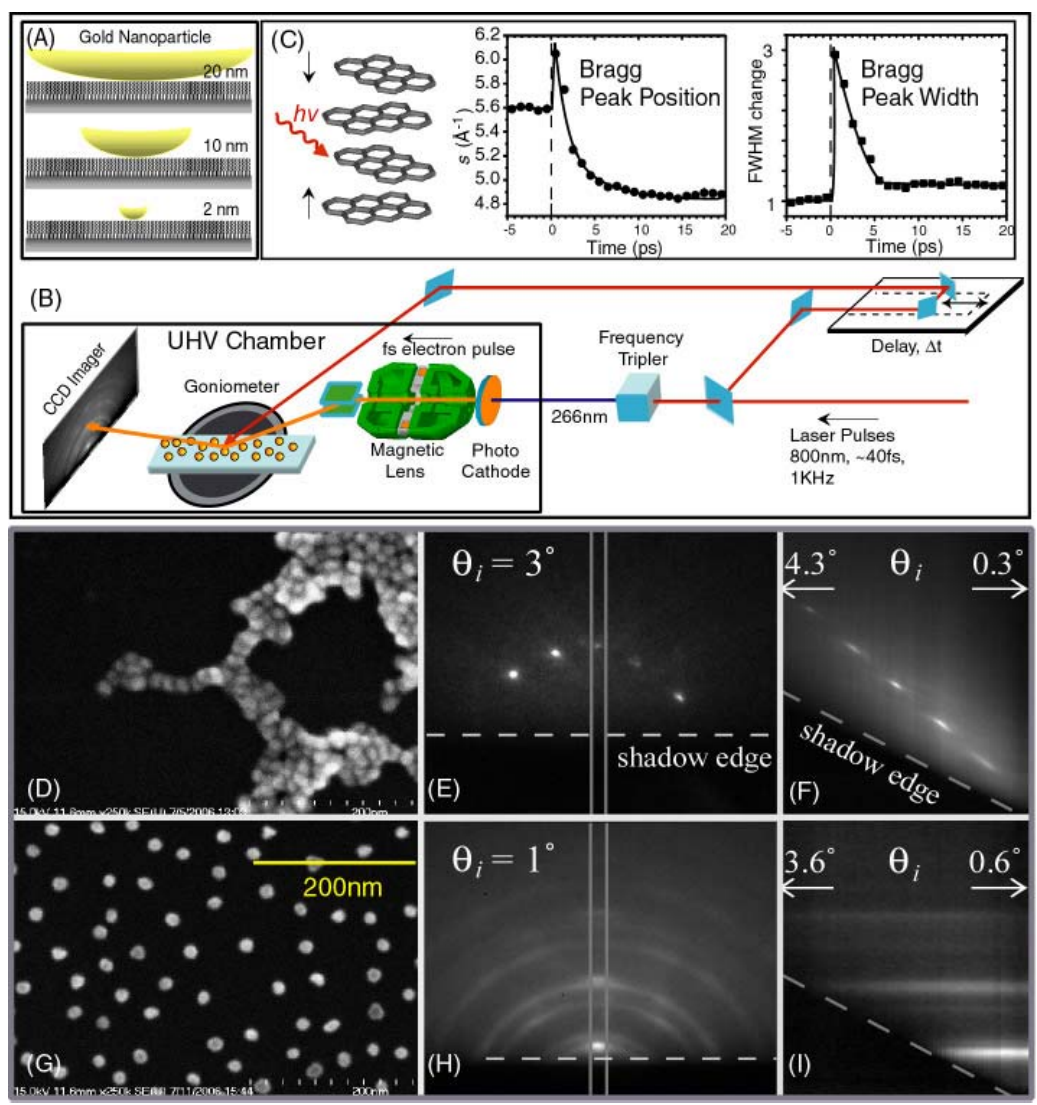

**Figure 1. (A)** Au nanoparticles dispersed on a self-assembled molecular interface **(B)** Pump-probe arrangement of UEC. **(C)** Zero-of-time determination using the diffraction signals from the photomechanical responses of graphite multilayers. **(D)** SEM image of 20 nm Au NPs dispersed on the surface without proper buffering. **(E)** Diffraction pattern from (D) showing Bragg spots of silicon substrate. **(F)** Rocking curve analysis gated at central streak in (E) with varying incident angle $\theta i$. **(G)** SEM image of 20 nm Au NPs with proper buffering. **(H)** Diffraction pattern from (G) showing Debye-Scherrer diffraction rings and Bragg spots from buffer layer (Si, N and C stack layers in self-assembled aminosilane, spacing 2.2Å, tilt angle 31°). **(I)** Rocking curve of (H).

**Figure 2**

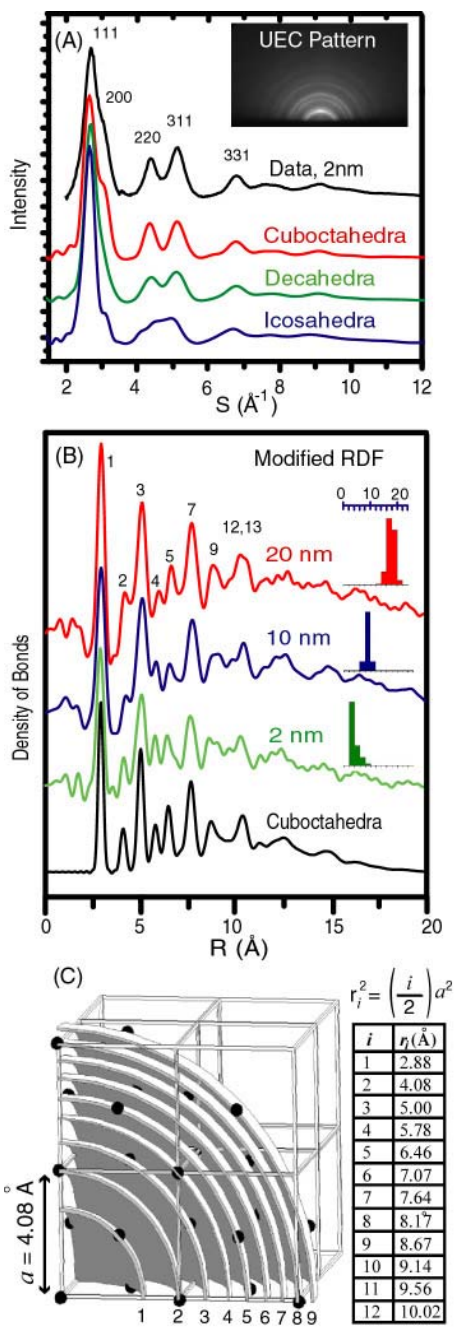

**Figure 2.** Structure analyses of size-selected Au nanoparticles. **(A)** 1D diffraction intensity curve (black) obtained from radial averaging of the Debye-Scherrer UEC pattern (inset) of the surface supported 2nm NPs. Also shown are simulations generated from 2nm structures (cuboctahedra, decahedra, and icosahedra) of Au NPs at 300 K. The indices show associated Bragg reflection planes based on an FCC structure. **(B)** Experimental modified radial distribution functions (mRDFs) of static Au NPs along with theoretical prediction for cuboctahedra. The numeric labels represent the bond order in the FCC distance table (below). **(C)** FCC coordination shells corresponding to interatomic distances $r_i$, calculated based on the bond order $i$ and the Au lattice constant a=4.08 Å.

**Figure 3**

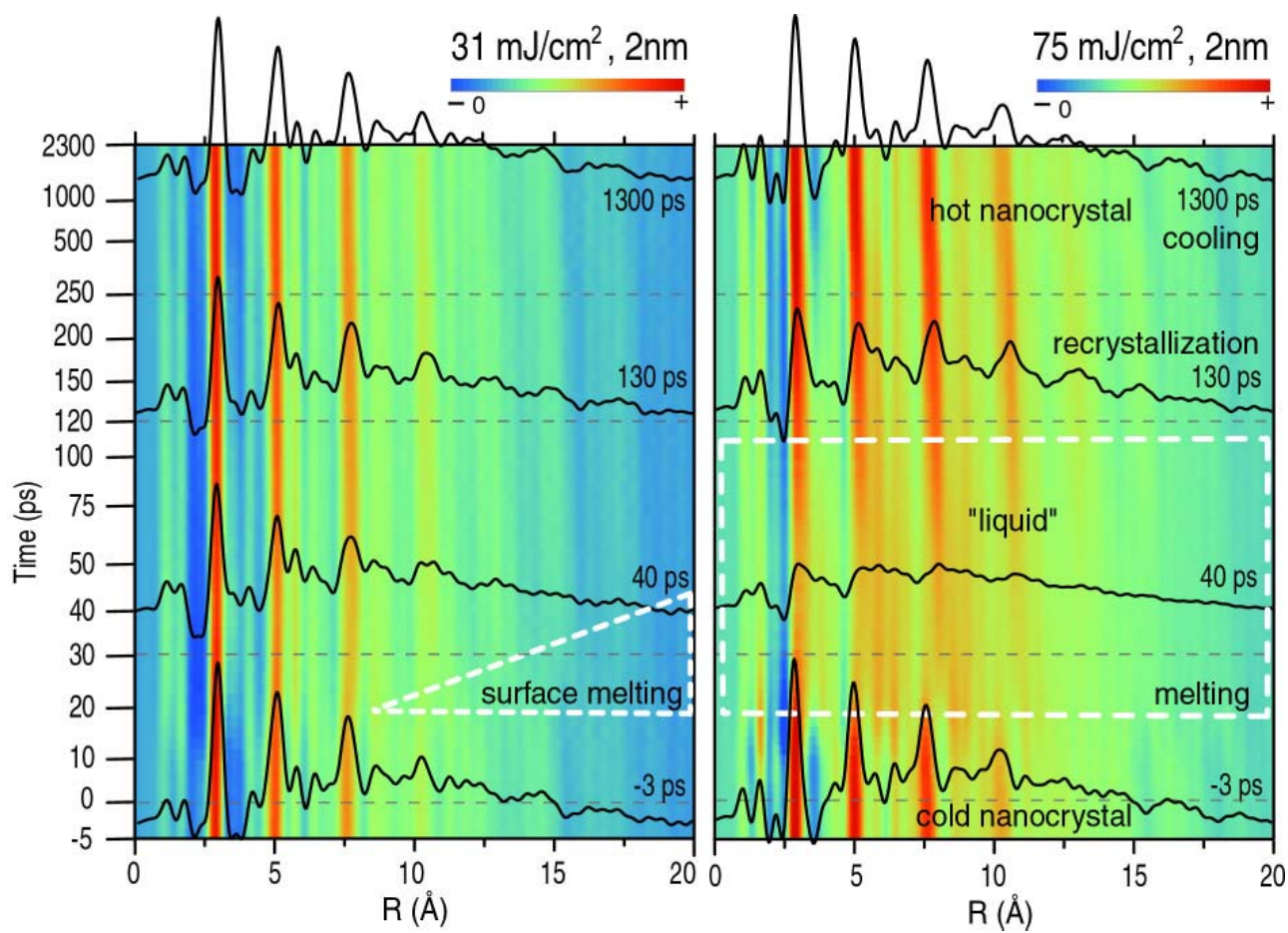

**Figure 3.** The melting dynamics of 2nm Au nanoparticles. **(Left)** mRDF map constructed by stacking mRDFs of UEC patterns at a sequence of delays between 5-2300 ps at irradiation fluence F=31 mJ/cm$^2$. Surface melting (enclosed by the dashed white line) is visible. **(Right)** mRDF map for F=75 mJ/cm$^2$. Full scale melting is observed. The liquid state (enclosed by dashed white line) is characterized by the drop of 2$^{nd}$ nearest density (at ~ 5Å) to *(1-1/e)* of the static value (at negative time).

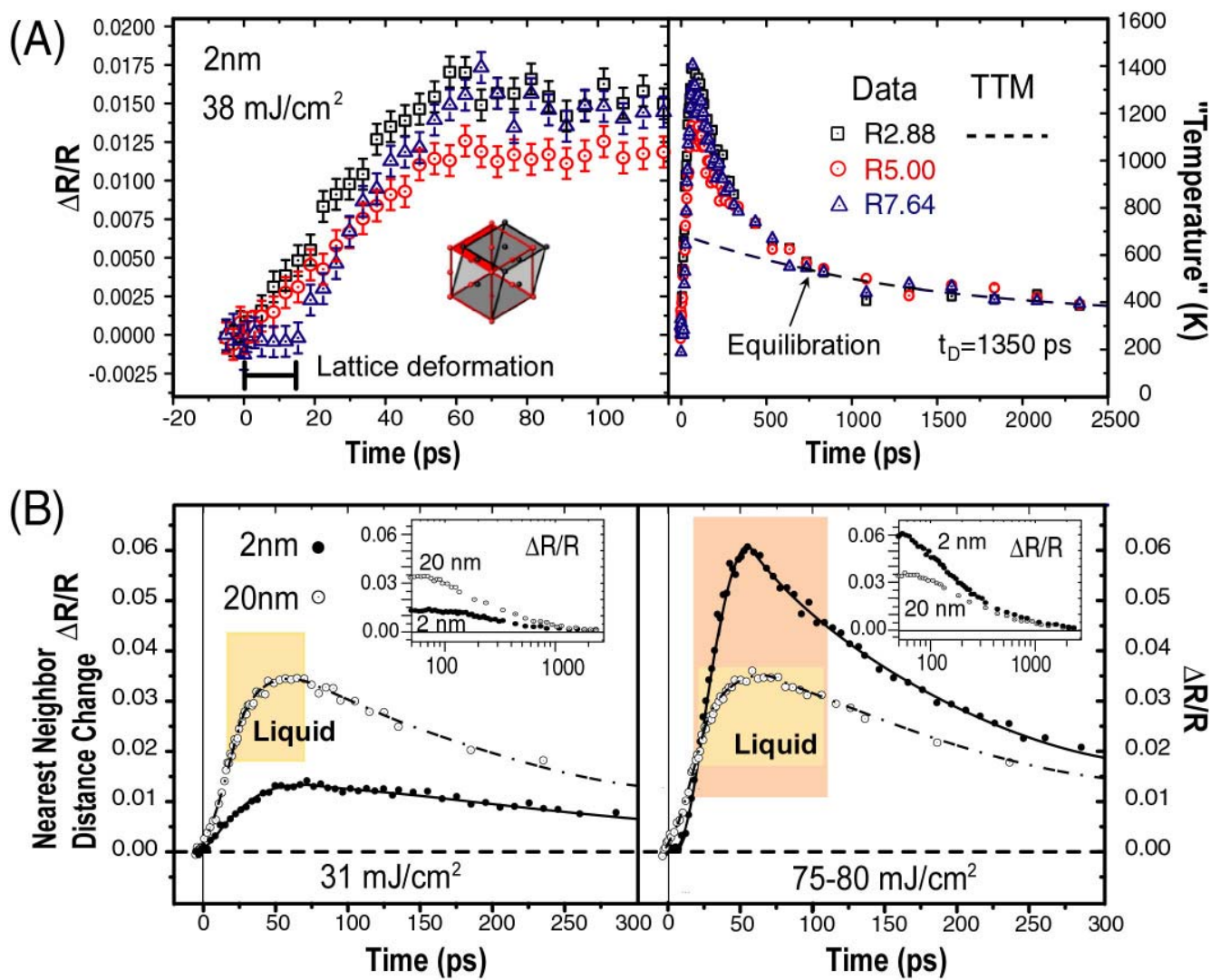

**Figure 4. (A)** The left panel shows the short-time relative distance change *ΔR/R* of dominant mRDF peaks (numbered 1, 3, and 7 in Fig. 2B) for 2nm Au NPs at fluence of 38 mJ/cm$^2$, from which the lattice deformation can be deduced. The right panel shows the extension of these bonds for longer times and the corresponding local temperature deduced from the bond extensions (see text) compared with a TTM calculation. $t_D$ is the thermal relaxation time to the environment obtained by fitting data to a two-temperature model (TTM) after equilibration. **(B)** Temporal evolution of the relative distance change *ΔR/R* of nearest neighbor bond (~2.88Å) in 2nm (solid line and symbol) and 20nm (dash-dash-dot line and open symbol) NPs, irradiated under F=31 mJ/cm$^2$ in the left panel, and F = 75 (for 2nm NPs) and 80 (for 20nm NPs) mJ/cm$^2$ in the right panel. The insets show the corresponding *ΔR/R* dynamics at long times (50-2750 ps).

16